\pdfoutput=1

\documentclass[3p,times]{elsarticle}

\usepackage{ecrc}

\usepackage[breaklinks]{hyperref}
\usepackage[utf8x]{inputenc}
\usepackage{subfig}
\usepackage{amsmath}
\usepackage{graphicx}

\volume{00}

\firstpage{1}

\journalname{Nuclear Physics A}

\runauth{T. Lappi and H. Mäntysaari}

\jid{nupha}

\jnltitlelogo{Nuclear Physics A}

\usepackage{amssymb}

\biboptions{comma,square,sort&compress}

\usepackage[figuresright]{rotating}

\newcommand{\der}{\mathrm{d}}
\newcommand{\rt}{{\mathbf{r}_T}}

\newcommand{\bt}{{\mathbf{b}_T}}
\newcommand{\st}{{\mathbf{s}_T}}

\newcommand{\pt}{{\mathbf{p}_T}}
\newcommand{\qt}{{\mathbf{q}_T}}
\newcommand{\kt}{{\mathbf{k}_T}}

\newcommand{\cf}{C_\mathrm{F}}

\newcommand{\qs}{Q_\mathrm{s}}

\newcommand{\qso}{Q_\mathrm{s0}}

\newcommand{\lqcd}{\Lambda_{\mathrm{QCD}}}
\newcommand{\as}{\alpha_{\mathrm{s}}}

\newcommand{\Ncal}{{\mathcal{N}}}

\begin{document}



\hypersetup{pdfauthor={T. Lappi and H. M\"antysaari},pdftitle={Particle production from the Color Glass Condensate: proton-nucleus collisions in light of the HERA data}}



\dochead{}


\author[jyfl,hip]{T. Lappi}

\author[jyfl]{H. M\"antysaari}

\address[jyfl]{
Department of Physics, %
 P.O. Box 35, 40014 University of Jyv\"askyl\"a, Finland
}
\address[hip]{
Helsinki Institute of Physics, P.O. Box 64, 00014 University of Helsinki,
Finland
}

\title{
Particle production from the Color Glass Condensate: proton-nucleus collisions in light of the HERA data
}

\begin{abstract}

We compute single inclusive hadron production in proton-proton and proton-nucleus collisions consistently within the CGC framework. The parameters in the calculations are obtained from electron-proton DIS and standard nuclear geometry. We obtain a good description of the DIS data without an anomalous dimension in the initial condition of the BK evolution and get a good agreement with the available single inclusive proton-proton and proton-nucleus data.

\end{abstract}

\maketitle

\section{Introduction}

The proton-nucleus run at the CERN Large Hadron Collider has produced many interesting results about particle correlations~\cite{Abelev:2012ola}, multiplicities~\cite{ALICE:2012xs} and nuclear suppression in charged particle spectra~\cite{ALICE:2012mj}. As it is expected that a significant amount of quark gluon plasma (QGP) is not created in these collisions, it is possible to study initial state cold nuclear matter effects by analysing the proton-nucleus data. Good understanding of the initial state physics is crucial for the correct interpretation of the heavy ion results where the properties of the QGP are studied.

In collider experiments at high energy the structure of the scattering hadron (proton or nucleus) is probed at small Bjorken-$x$. In this region the gluon densities are large and non-linear phenomena, such as gluon recombination, become important. A convenient way to describe 
these effects is provided by the Color Glass Condensate (CGC)  effective field theory that has been shown to agree well with the available small-$x$ data, see e.g. Refs. \cite{Tribedy:2011aa,Albacete:2012xq,Lappi:2012nh, Lappi:2013am}. For a recent review of the CGC phenomenology, we refer the reader to Ref. \cite{Albacete:2014fwa}. Because the gluon densities scale as $A^{1/3}$, the non-linear phenomena are 
enhanced when the target is changed from a proton to a heavy nucleus.
The p+Pb run at the LHC allows us to probe the non-linearly behaving QCD matter
in a kinematical region never explored so far.

\section{Deep inelastic scattering baseline}

The structure of a hadron can be studied accurately in deep inelastic 
scattering (DIS) where a (virtual) photon scatters off the hadron. Precise
measurements of the proton structure at HERA have been a crucial test for the 
CGC, and recent analyses have confirmed that the CGC
description is consistent with all the available small-$x$ structure function data, see e.g. Ref.~ 
\cite{Albacete:2010sy}. In this work we compute, as explained in more detail in Ref. \cite{Lappi:2013zma}, 
single inclusive hadron production in proton-proton and proton-nucleus collisions consistently within the CGC
framework. As an input we use only the HERA data for the inclusive DIS cross section and standard nuclear geometry.


The H1 and ZEUS collaborations have measured the proton
structure functions $F_2$ and $F_L$ and published very precise combined
results for the reduced cross section $\sigma_r$~\cite{Aaron:2009aa}.
The structure
functions are related to the virtual photon-proton cross sections 
$\sigma_{T,L}^{\gamma^*p}$ for transverse (T) and longitudinal (L) photons, computed as
\begin{equation}\label{eq:sigmagammastar}
	\sigma_{T,L}^{\gamma^*p}(x,Q^2) = 2\sum_f \int \der z \int \der^2 \bt |\Psi_{T,L}^{\gamma^* \to f\bar f}|^2 \Ncal(\bt, \rt, x).
\end{equation}
Here $\Psi_{T,L}^{\gamma^* \to f\bar f}$ is the photon light cone wave 
function describing how the photon fluctuates to a quark-antiquark pair,
computed from light cone QED.

The Bjorken-$x$, or equivalently energy evolution of the dipole-proton amplitude $\Ncal$ is given by the BK 
equation, for which we use the running
coupling corrections derived in Ref.~\cite{Balitsky:2006wa}. The initial condition for the dipole amplitude is a non-perturbative input, for which we use a modified McLerran-Venugopalan model~\cite{McLerran:1994ni}

\begin{table}
\begin{center}
\begin{tabular}{|l||r|r|r|r|r|r|r|}
\hline
Model & $\chi^2/\text{d.o.f}$ &  $\qso^2$ [GeV$^2$] & $\qs^2$ [GeV$^2$] & $\gamma$ & $C^2$ & $e_c$ & $\sigma_0/2$ [mb] \\
\hline\hline
MV & 2.76 & 0.104 & 0.139  & 1 & 14.5 & 1 & 18.81 \\
MV$^\gamma$ & 1.17 & 0.165 & 0.245 & 1.135 & 6.35 & 1 & 16.45 \\
MV$^e$ & 1.15 & 0.060 & 0.238  & 1 & 7.2 & 18.9 & 16.36 \\
\hline
\end{tabular}
\caption{Parameters from fits to HERA reduced cross section data at $x<10^{-2}$ and $Q^2<50\,\mathrm{GeV}^2$ for different initial conditions. Also the corresponding initial saturation scales $\qs^2$ defined via equation $\Ncal(r^2=2/\qs^2)=1-e^{-1/2}$ are shown. The parameters for the MV$^\gamma$ initial condition are obtained by the AAMQS collaboration \cite{Albacete:2010sy}.
}
\label{tab:params}
\end{center}
\end{table}

\begin{equation}
\label{eq:aamqs-n}
	\Ncal(\rt) = 1 - \exp \left[ -\frac{(\rt^2 \qso^2)^\gamma}{4} \ln \left(\frac{1}{|\rt| \lqcd}+e_c \cdot e\right)\right],
\end{equation}
where we have generalized the AAMQS~\cite{Albacete:2010sy} form by also
allowing the constant inside the logarithm, which plays a role of an infrared cutoff, to be different from $e$. The other fit parameters are the
anomalous dimension $\gamma$, the initial saturation scale $\qso^2$, proton transverse area $\sigma_0/2$ and the scaling factor $C^2$ of the QCD scale $\lqcd$  in the expression of $\as$, see Ref. \cite{Lappi:2013zma}.

The unknown parameters are obtained by performing a fit to the HERA combined structure function data. We consider the standard MV model, where we set $\gamma=1$ and $e_c=1$, and compare with the MV$^\gamma$ model (where $\gamma$ is a free parameter, fitted in Ref.~\cite{Albacete:2010sy}) and the MV$^e$ parametrization where $\gamma=1$ but $e_c$ is free. The two modified parametrizations give a much better fit to the HERA $\sigma_r$ data, see Table \ref{tab:params}.

\section{Single inclusive hadron production in CGC}
\label{sec:sinc}

The gluon spectrum in hadron collisions can be obtained
by solving the classical Yang-Mills equations of
motion for the color fields. When computing the production of the high-$|\pt|$ hadrons it has been shown numerically~\cite{Blaizot:2010kh}
that this solution is well approximated by the 
 $k_T$-factorized formula ~\cite{Kovchegov:2001sc}
\begin{equation}
\label{eq:ktfact-bdep}
\frac{\der \sigma}{\der y \der^2 \pt \der^2 \bt} = \frac{2 \as}{\cf \kt^2 } \int \der^2 \qt \der^2 \st \frac{\varphi_p(\qt,\st)}{\qt^2} 
	 \frac{\varphi_p(\pt-\qt,\bt-\st)}{(\pt-\qt)^2},
\end{equation}
where $\varphi_p$ is the dipole unintegrated gluon
distribution (UGD) of the 
hadron and $\bt$ is the impact parameter. In collisions where $|\pt|$ is much larger than the saturation scale of one of the colliding objects (e.g. in proton-proton collision at forward rapidity) one can derive the hybrid formalism result
\begin{equation}
	\label{eq:pp-hybrid}
	\frac{\der N}{\der y \der^2 \pt} = \frac{\sigma_0/2}{\sigma_\text{inel}} \frac{1}{(2\pi)^2} xg(x,\pt^2) \tilde S^p(\pt),
\end{equation}
where $xg$ is the standard collinear factorization gluon distribution function satisfying the DGLAP equation and $\tilde S(\pt)$ is the Fourier transform of $1-\Ncal$.

\begin{figure}
\begin{minipage}[t]{0.48\linewidth}
\centering
\includegraphics[width=1.05\textwidth]{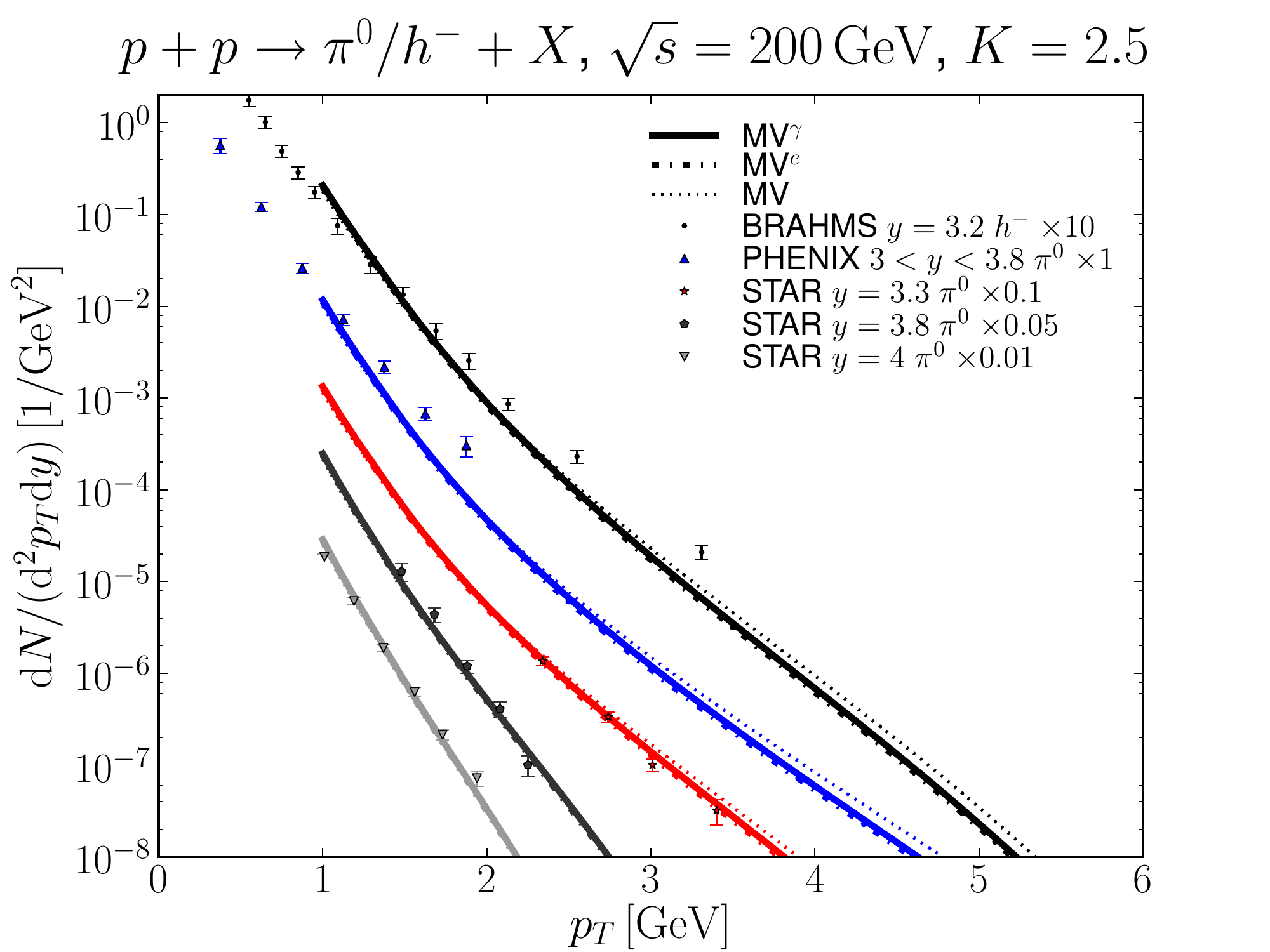} 
\caption{
Single inclusive $\pi^0$ and negative hadron production computed using MV, MV$^e$ and MV$^\gamma$ initial conditions compared with RHIC data from STAR~\cite{Adams:2006uz}, PHENIX~\cite{Adare:2011sc} and BRAHMS~\cite{Arsene:2004ux} collaborations. \\
}\label{fig:rhic_pp_yield}
\end{minipage}
\hspace{0.5cm}
\begin{minipage}[t]{0.48\linewidth}
\centering
\includegraphics[width=1.05\textwidth]{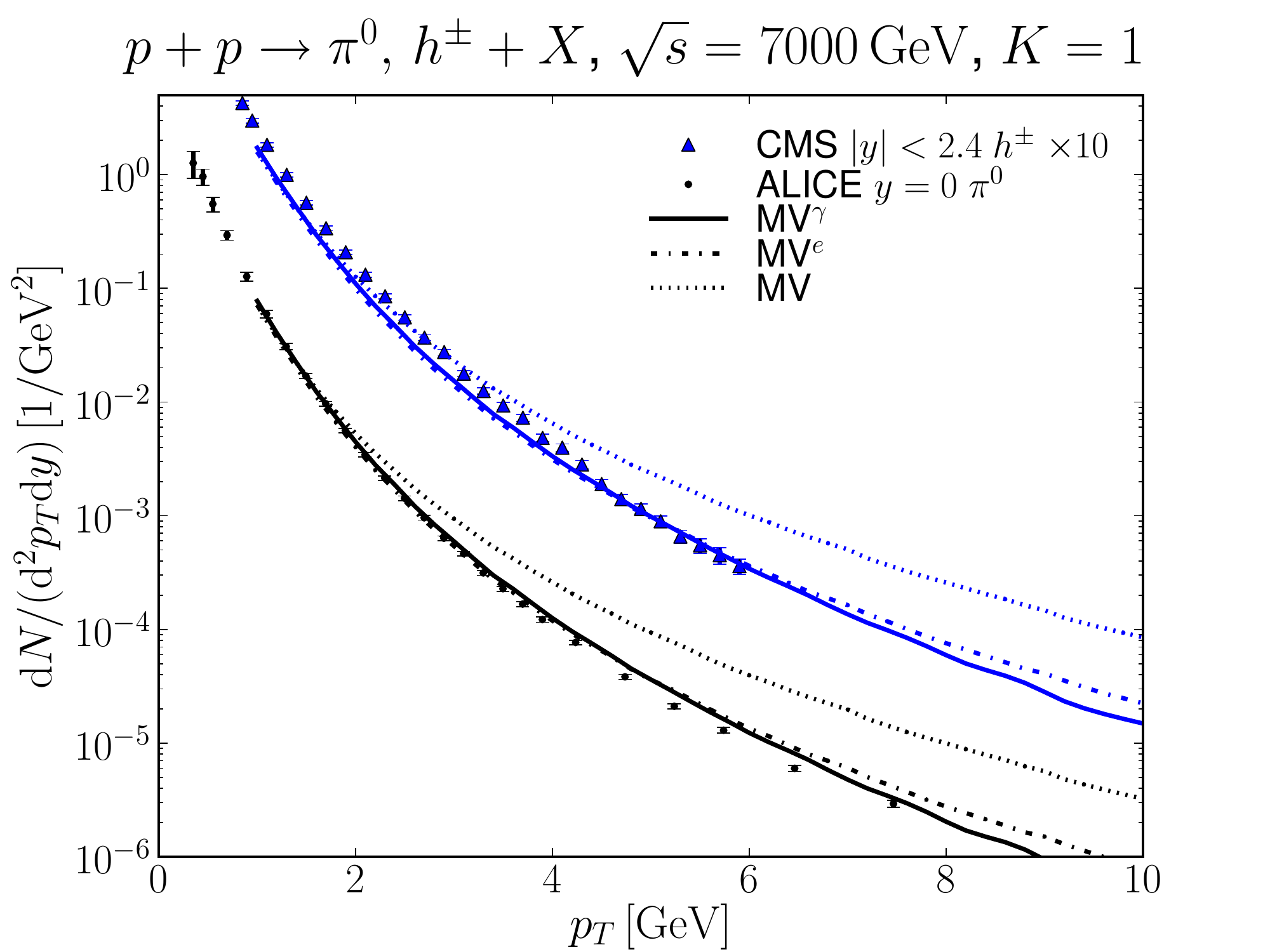} 
\caption{
Single inclusive $\pi^0$ production computed using MV, MV$^\gamma$ and MV$^e$ initial conditions at $\sqrt{s}=7000$ GeV compared with ALICE $\pi^0$ \cite{Abelev:2012cn} and CMS charged hadron data~\cite{Khachatryan:2010us}.\\
}\label{fig:lhc_pp_yield}
\end{minipage}
\end{figure}

In Fig. \ref{fig:rhic_pp_yield} we show the single inclusive pion and charged hadron production yields and compare our results to the RHIC data. All initial conditions for the dipole amplitude give basically the same invarian yield, but a normalization factor $K\sim 2.5$ is required in order to match the data. As we have consistently used different proton areas (inelastic proton-proton cross section $\sigma_\text{inel}$ and proton size in $\gamma^*p$ collisions $\sigma_0/2$) in the calculations, the $K$ factor tells how much the LO CGC result differs from the data. Comparison with the midrapidity LHC data is shown in Fig. \ref{fig:lhc_pp_yield}. Now the standard MV model clearly disagrees with the $p_T$ slope, and MV$^\gamma$ and MV$^e$ results are basically identical. No $K$ factor is required when comparing with the LHC data.

\section{Proton-nucleus collisions}

In order to describe proton-nucleus collisions we need the dipole-nucleus amplitude $\Ncal^A$. Due to the lack of small-$x$ DIS data, we can not perform a similar fit as with the proton targets. Instead we generalize the dipole-proton amplitude $\Ncal$ for nuclei using the optical Glauber model and write the initial condition for the BK evolution as
\begin{equation}
	\Ncal^A(\rt,\bt) = 1-\exp \left(-\frac{AT_A(\bt)}{2} \sigma_\text{dip}^p \right),
\end{equation}
where $\sigma_\text{dip}^p=\sigma_0 \Ncal(\rt)$ is the total dipole-proton cross section and $T_A$ is the Woods-Saxon distribution. In order to satisfy $\Ncal^A(\rt,\bt)\to 1$ at large dipoles we use here a linearized version of the dipole amplitude $\Ncal$ from Eq. \eqref{eq:aamqs-n}. 

The BK evolution would cause the dipole amplitude to evolve rapidly at the dilute edges of the nucleus causing an unphysical growth of the nuclear size. In order to obtain a reliable estimate for the contribution to the particle production from the edges of the nucleus we approximate the differential yield from the edges as $\der N^{pA}=N_\text{bin}\der N^{pp}$,
which is equivalent to imposing $R_{pA}=1$ at large impact parameters.

\begin{figure}[tb]
\begin{minipage}[t]{0.48\linewidth}
\centering
\includegraphics[width=1.05\textwidth]{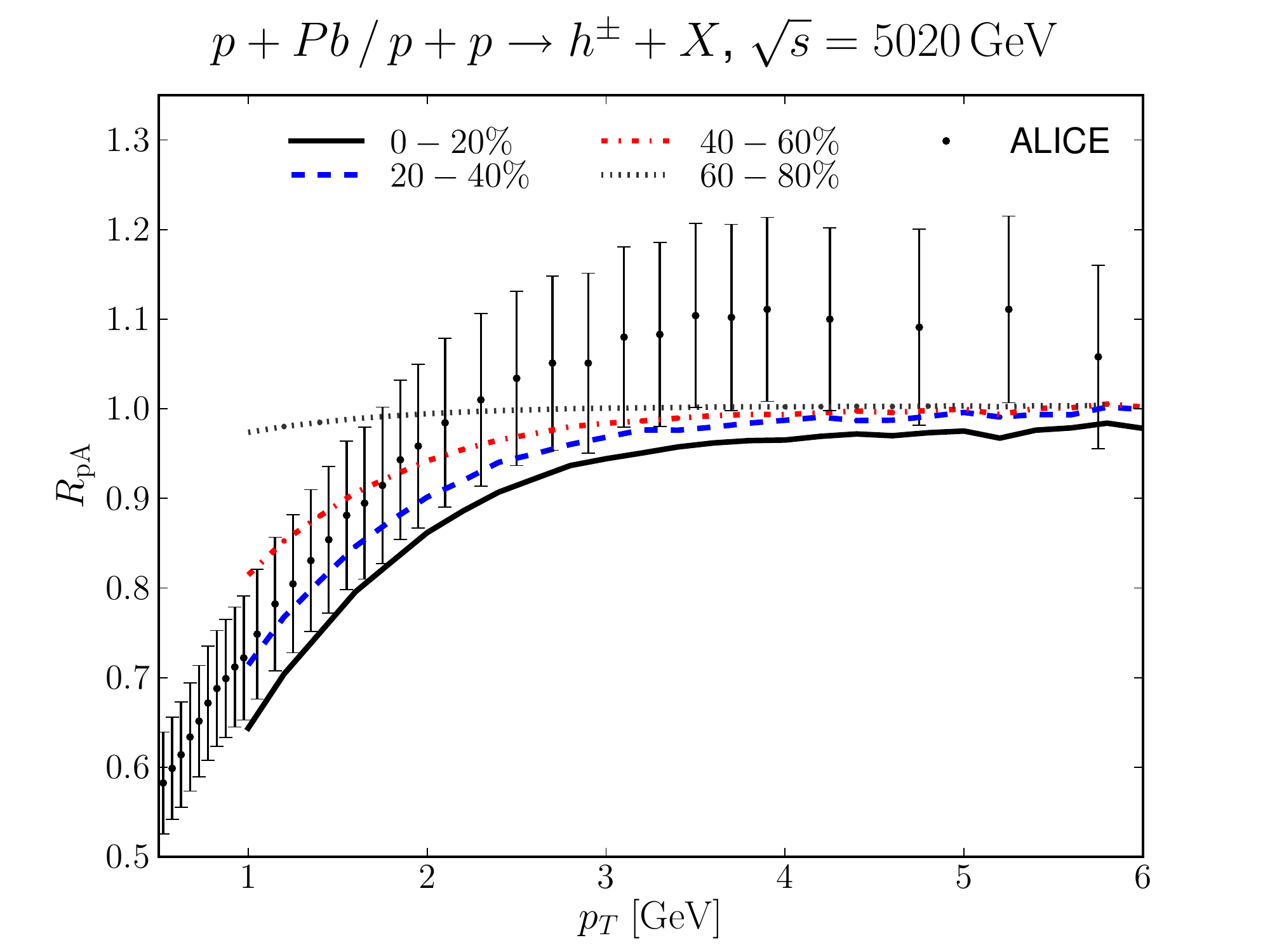}
\caption{
Centrality dependence of the nuclear modification factor $R_{pA}$ at $\sqrt{s}=5020$ GeV p+Pb collisions compared with the minimum bias ALICE data~\cite{ALICE:2012mj}.
}\label{fig:rpa_y0_c}
\end{minipage}
\hspace{0.5cm}
\begin{minipage}[t]{0.48\linewidth}
\centering
\includegraphics[width=1.05\textwidth]{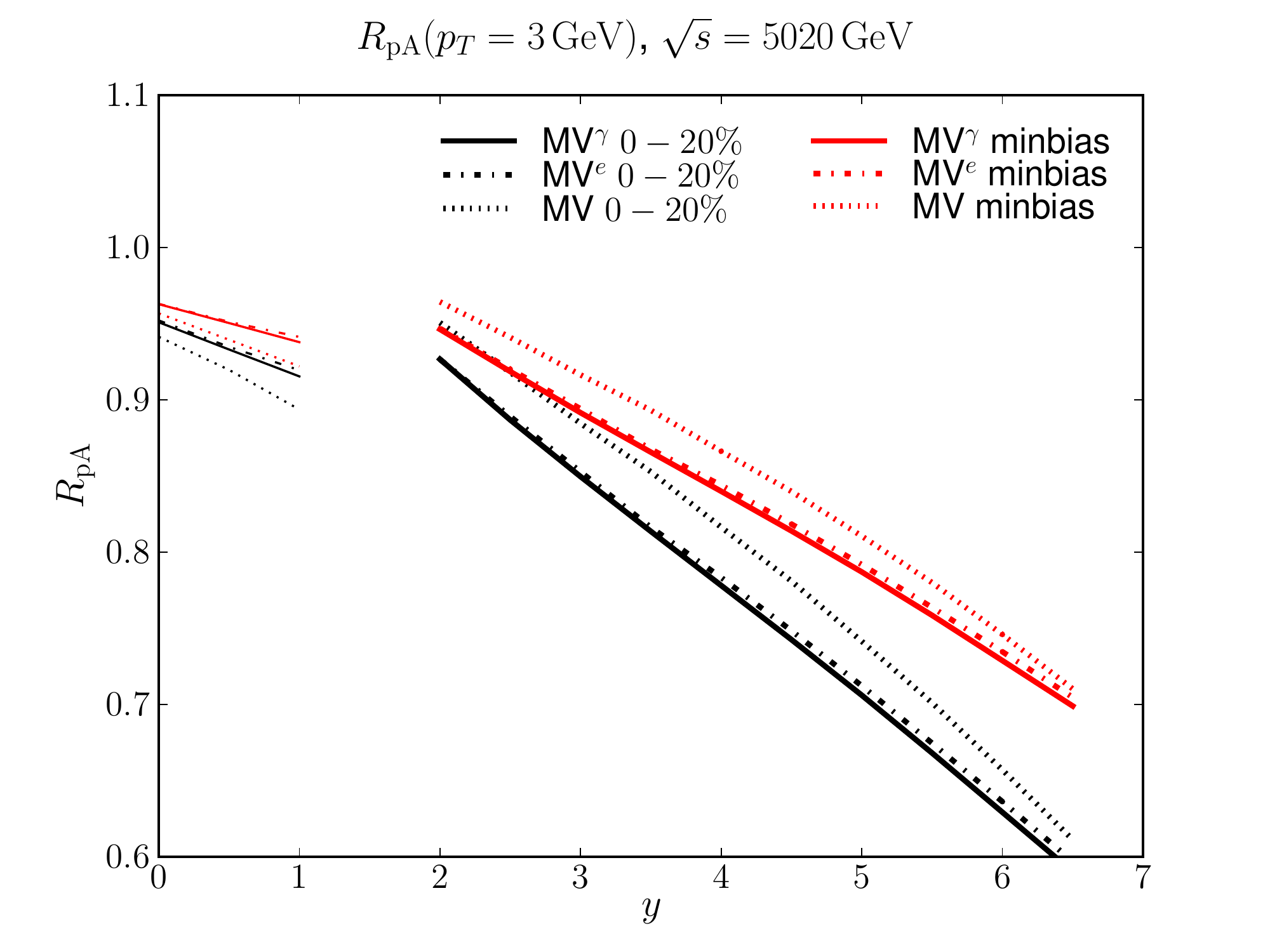}
\caption{
Rapidity dependence of the nuclear modification factor $R_{pA}$ for $|\pt|=3$ GeV neutral pion production at most central and minimum bias collisions.
}\label{fig:rpa_fixedpt}
\end{minipage}
\end{figure}

In Fig. \ref{fig:rpa_y0_c} we show the nuclear suppression factor $R_{pA}$ computed at different centrality classes and compare with the minimum bias ALICE data~\cite{ALICE:2012mj}. We get explicitly $R_{pA}\to 1$ at midrapidity and large transverse momentum, and our results are consistent with the ALICE data. In Fig. \ref{fig:rpa_fixedpt} we show the $R_{pA}$ for $3\,\mathrm{GeV}$ neutral pions as a function of rapidity. Close to midrapidity we use the $\kt$ factorization from Eq. \eqref{eq:ktfact-bdep}, and the hybrid formalism is used at forward rapidities. The evolution speed is given by the BK equation, and thus the rapidity evolution of $R_{pA}$ is a solid prediction of the CGC.


\section*{Acknowledgements}
This work has been supported by the Academy of Finland, projects 133005, 
267321, 273464 and by computing resources from CSC -- IT Center for 
Science in Espoo, Finland. H.M. is supported by the Graduate School of 
Particle and Nuclear Physics.

\bibliography{../../refs}
\bibliographystyle{h-physrev4mod2}

\end{document}